\documentclass[aip,pre,twocolumn,superscriptaddress,floatfix,10pt]{revtex4}
\usepackage{graphicx}
\usepackage{amsmath}
\usepackage{color}
\begin{document}

\title{Tuning morphology and thermal transport of asymmetric smart polymer blends by macromolecular engineering}

\author{Daniel Bruns}
\affiliation{Stewart Blusson Quantum Matter Institute, University of British Columbia, Vancouver BC V6T 1Z4, Canada}
\affiliation{Department of Physics and Astronomy, University of British Columbia, Vancouver BC V6T 1Z1, Canada}
\author{Tiago Espinosa de Oliveira}
\affiliation{Universidade Federal do Rio Grande do Sul, Porto Alegre, Brazil}
\author{J\"org Rottler}
\affiliation{Stewart Blusson Quantum Matter Institute, University of British Columbia, Vancouver BC V6T 1Z4, Canada}
\affiliation{Department of Physics and Astronomy, University of British Columbia, Vancouver BC V6T 1Z1, Canada}
\author{Debashish Mukherji}
\email[]{debashish.mukherji@ubc.ca}
\affiliation{Stewart Blusson Quantum Matter Institute, University of British Columbia, Vancouver BC V6T 1Z4, Canada}

\date{\today}

\begin{abstract}
	A grand challenge in designing polymeric materials is to tune their
properties by macromolecular engineering. In this context, one of the drawbacks that often limits
broader applications under high temperature conditions is their poor thermal conductivity $\kappa$.
Using molecular dynamics simulations, we establish a structure-property relationship
in hydrogen bonded polymer blends for possible improvement of $\kappa$. For this purpose, we investigate two experimentally relevant hydrogen bonded systems$-$
one system consists of short poly({\it N}-acryloyl piperidine) (PAP) blended with longer
chains of poly(acrylic acid) (PAA) and the second system is a mixture of
PAA and short poly(acrylamide) (PAM) chains.
Simulation results show that PAA-PAP blends are at the onset of phase separation over the full range of PAP monomer mole fraction
$\phi_{\rm PAP}$, which intensifies even more for $\phi_{\rm PAP} > 0.3$. While PAA and PAP interact with preferential hydrogen bonding,
phase separation is triggered by the dominant van der Waals attraction between the hydrophobic side groups
of PAP. However, if PAP is replaced with PAM, which has a similar chemical structure as PAP without the
hydrophobic side group, PAA-PAM blends show much improved solubility. Better solubility
is due to the preferential hydrogen bonding between PAA and PAM. As a result, PAM
oligomers act as cross-linking bridges between PAA chains resulting in a three dimensional highly cross-linked network.
While $\kappa$ for PAA-PAP blends remain almost invariant with $\phi_{\rm PAP}$, PAA-PAM systems show improved $\kappa$
with increasing PAM concentration and also with respect to PAA-PAP blends. Consistent with the
theoretical prediction for the thermal transport of amorphous polymers, we show
that $\kappa$ is proportional to the materials stiffness, i.e., the bulk modulus $K$ and sound velocity $v$
of PAA-PAM blends. However, no functional dependence between $\kappa$ and $K$ (or $v$) is observed for the
immiscible PAA-PAP blends.
\end{abstract}

\maketitle

\section{Introduction}
\label{sec:intro}

Polymers are ubiquitous in our everyday life, finding a variety of applications ranging from physics to materials science
and chemistry to biology \cite{cohen10natmat,mukherji14natcom,sissi14natcom,winnik15review,hoogenboom,mukherji17natcom}.
The properties of polymers are intimately linked to large conformational and compositional fluctuations. Because of the
molecular flexibility, polymer conformations can be tuned almost {\it at will} for desired applications
and thus provide a robust platform for advanced functional materials design. However, one of the drawbacks of
standard commodity polymeric materials is the poor thermal conductivity $\kappa$ in their amorphous states
\cite{choy77pol,shen10natnano,cahill11prb}. This is partially because of rather weak van der Waals interactions dictating polymer properties.
Therefore, it is desirable to tune $\kappa$ of polymeric materials, especially when they are used in high temperature
environments.

One of the standard protocols to improve $\kappa$ of polymeric materials is blending them with
materials having $\kappa$ values exceeding the thermal conductivity of metals, such as carbon
based materials \cite{keb11jap,davide13prb,kodama17nm,mahoney16pol}. In this context, following the arguments of continuum theory,
one should expect to increase $\kappa$ of polymer composites with increasing concentration of the high $\kappa$ guest.
Moreover, establishing a tunable structure-function relationship in these composite materials is
often difficult because they exhibit large spatial and temporal heterogeneities.
Furthermore, a significant improvement
in $\kappa$ also requires concentrations of external guest material exceeding their percolation threshold,
thus also losing the inherent property and flexibility of the host polymeric systems.
Therefore, a more attractive protocol to tune $\kappa$ is to strengthen
microscopic interactions within the polymer system itself.
Here, smart polymers may serve as ideal candidates.

\begin{figure}[ptb]
\includegraphics[width=0.46\textwidth,angle=0]{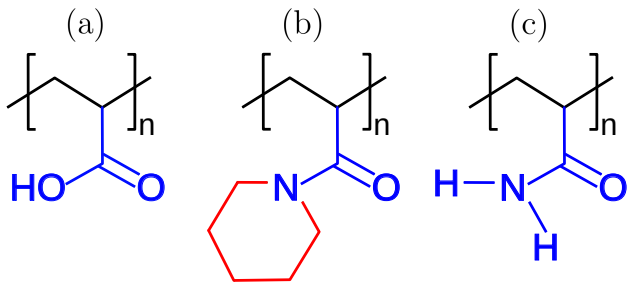}
\caption{Schematic representations of$-$ (a) poly(acrylic acid) (PAA),
        (b) poly({\it N}-acryloyl piperidine) (PAP) and (c) poly(acrylamide) (PAM) systems.
The length of PAA was chosen as $N_{\ell}^{\rm PAA} = 30$, the lengths of PAP and PAM is taken as $3$.
Highlighted blue regions are possible groups forming hydrogen bonding, while the black and red regions are
hydrophobic in nature.
\label{fig:schem}}
\end{figure}

A polymer is referred to as ``smart", when they exhibit fast responsiveness to a change in their environment
in solutions and are typically dictated by hydrogen bonding whose strength typically falls within the range of 4-8 $k_{\rm_B}T$, thus
exceeding significantly the van der Waals pair interactions that are only of the order of less than $k_{\rm_B}T$
\cite{cohen10natmat,mukherji14natcom,winnik15review,hoogenboom}.
Therefore, the solvent-free dry states of these hydrogen bonded polymers are dictated by strong inter-polymer
interactions. In this context, there is considerable interest to study thermal transport in
tunable polymer materials \cite{pipe14nm,cahill16mac,kim17sci}, smart hydrogels \cite{tang17pol,zu18acsapp}, concentrated polymer
solutions \cite{li18maclett}, and solid polyelectrolytes \cite{arxiv}. In particular, a recent work uses the idea of hydrogen bonding
as a tunable interaction to propose a wide range of polymer blends, where a significant increase in $\kappa$ was observed \cite{pipe14nm}.
One of the interesting systems this experiment proposes is an asymmetric blend of poly({\it N}-acryloyl piperidine) (PAP)
and poly(acrylic acid) (PAA) with M$_{\rm w}^{\rm PAA} \gg$ M$_{\rm w}^{\rm PAP}$, where M$_{\rm w}$ is the molecular weight.
Schematic representations of chemical structures of PAA and PAP are shown in Figures \ref{fig:schem}(a-b).
It was reported that, for a PAP monomer mole fraction $\phi_{\rm PAP} \sim 0.3$,
$\kappa$ increases by a factor of about 6$-$7 times with respect to $\kappa \sim 0.25$ Wm$^{-1}$K$^{-1}$ of pure PAA (or pure PAP) \cite{pipe14nm}.
This increase was attributed to the strong PAA-PAP hydrogen bonding, which also demands as prerequisite that the binary
solution of PAA and PAP are fairly miscible. In a separate experimental study, however, it was found that PAA and PAP phase separate,
while no variation in $\kappa$ was observed over the full range of $\phi_{\rm PAP}$ \cite{cahill16mac}.

If a system is phase separated, it is expected to show reduced $\kappa$ because of the
weakened interfacial interaction between two phase separated regions. However,
because of the delicate interplay between the van der Waals and hydrogen bonding interactions in these systems,
it is not always straightforward to predict molecular level morphology and its connection to
thermal transport, which is the motivation behind this study.
To best of our knowledge, there is no theoretical/computational work addressing
polymer blends in their dry states for tunable $\kappa$. In this work, we present
molecular dynamics simulation results establishing a structure-property relationship in polymer blends.
For this purpose, we investigate two experimentally relevant systems$-$ one system is a simulation replica of the PAA-PAP
blend reported earlier \cite{pipe14nm} and the second system consists of a PAA and poly(acrylamide) (PAM)
blend, where PAM has a very similar chemical structure as PAP without the hydrophobic carbon ring, see
Figure \ref{fig:schem}(c).

The remainder of the paper is organized as follows: In section \ref{sec:method} we sketch our methodology.
Results and discussions are presented in sections \ref{sec:res} and finally
the conclusions are drawn in section \ref{sec:conc}.

\section{Method and model}
\label{sec:method}

In this work all-atom molecular dynamics simulations are performed at two stages: the GROMACS 4.6 package \cite{gro}
was used for the equilibration and structural analysis of the polymer blends, while the
LAMMPS package \cite{lammps} was used for the thermal transport calculations.

GROMACS simulations are performed in the NpT ensemble, where N is the number of particles in a system, p
is the isotropic pressure, and T is the temperature. T= 600 K is set for the initial simulations using
a Berendsen thermostat with a coupling constant 2 ps. This ensures that the polymer blends are in
their melt states. p is kept at 1 bar using a Berendsen barostat with a coupling time of 0.5 ps \cite{berend}.
Electrostatics are treated using the particle mesh Ewald method \cite{pme}. The interaction cutoff for non-bonded
interactions is chosen as 1.0 nm. The simulation time step is chosen as $\Delta t = 1$ fs and the equations
of motion are integrated using the leap-frog algorithm \cite{lfa}.
All bond vibrations are constrained using a LINCS algorithm \cite{linc}.

We investigate two different polymer systems, namely PAA-PAP and PAA-PAM blends.
Configurations consist of a total of 200 chains, where the length of PAA is chosen as $N_{\ell}^{\rm PAA} = 30$,
while the lengths of both PAP and PAM oligomers are taken as $N_{\ell}^{\rm PAP}$ (or $N_{\ell}^{\rm PAM}$) = 3.
In a series of simulations, the PAP monomer mole fractions $\phi_{\rm PAP}$ and PAM monomer
mole fractions $\phi_{\rm PAM}$ are varied between $0.0-1.0$,
where $\phi_i = 0.0$ corresponds to pure PAA and $\phi_i = 1.0$ represents pure PAP or PAM systems.
The standard OPLS force field was used for PAA and PAP systems \cite{OPLS96}. For PAM we have used modified parameters
that were developed by two of us earlier \cite{oliveira17jcp}. While both these force fields were previously used for
the study of polymers in solution, in the supplementary materials we provide evidence that these force fields are
also suitable to study dry polymer films.

After an initial equilibration of 20 ns, production runs are performed for 60 ns and the configurations were saved
each 50 fs for the calculation of the structural properties. These simulation runs are at least one order of magnitude larger than the
longest relaxation time $\tau \sim 5$ ns of a PAA chain with $N_{\ell}^{\rm PAA} = 30$ and T = 600 K,
which is estimated using the end-to-end distance auto-correlation function
$\left <{\bf R}(t)\cdot {\bf R}(0) \right> \sim e^{-t/\tau}$. During the production runs, observables such as
the radial distribution function ${\rm g}_{ij} (r)$ and the number of hydrogen bonds (h$-$bond) for different solution
components are calculated. H$-$bond are calculated using
the standard geometrical criterion implemented in GROMACS, i.e., a hydrogen bond exists when the
donor-acceptor distance is $\le0.35$ nm and the acceptor-donor-hydrogen angle is $\le 30^{\circ}$.

The final configurations from the GROMACS simulations were first quenched down to
$T = 300$ K and then imported in LAMMPS for $\kappa$ calculations.
Note that both pure PAA and PAP have their glass transition temperatures
$T_{\rm g} \sim 373$ K \cite{pipe14nm}, while $T_{\rm g}$ of bulk PAM exceeds 430 K \cite{tgpam}.
Therefore, without attempting to identify a precise $T_{\rm g}$ value for different blends and following a simple combination rule,
one expects $T = 300$ K to be well below their corresponding $T_{\rm g}$.
This, however, also assumes a priory that the bulk solution is miscible.

\begin{figure*}[ptb]
\includegraphics[width=0.91\textwidth,angle=0]{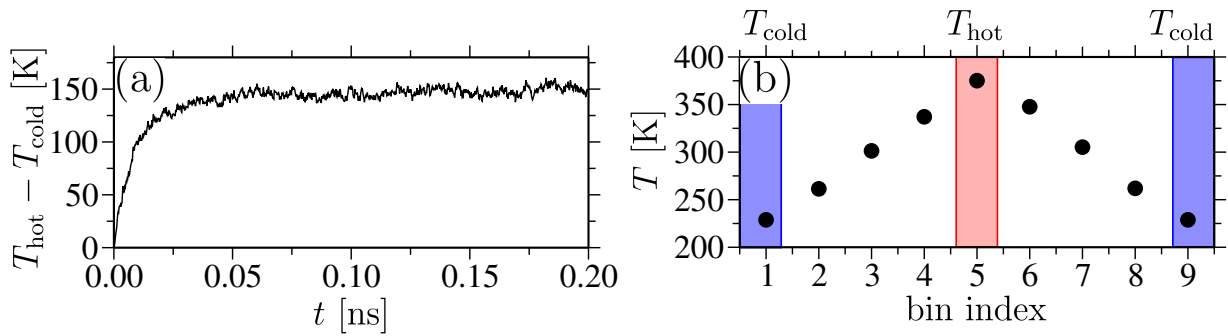}
        \caption{Part (a) shows the time equilibration of temperature difference between the hot $T_{\rm hot}$ and the cold $T_{\rm cold}$
        regions of a PAA-PAM blend for $\phi_{\rm PAM} = 0.286$. Part (b) shows the steady-state temperature profile with
        bin index along the $z-$axis of the simulation domain. Note that while we sub-divide simulation domain into
        eight chunks, nine chunks are shown because of the periodic boundary condition.
\label{fig:mp}}
\end{figure*}

For the calculation of $\kappa$, we employ a non-equilibrium method \cite{plathe}. In this method, a heat flux
$J$ through the system is generated over a simulation time $\mathcal{T}$ in the microcanonical ensemble by swapping atomic velocities between
a hot and a cold region of the simulation box,
\begin{equation}
        J=\frac {1}{2\,A\mathcal{T}} {\sum_{\text{swaps}}\Delta E_{\text{kin}}},
\end{equation}
where $\Delta E_{\text{kin }}$denotes the kinetic energy exchanged per swap, $A$ is the cross sectional area perpendicular to the direction
of heat flow, and the factor 2 accounts for the two directions of heat flow present in systems with periodic boundaries. As a result
of velocity swapping, once the system reaches its steady-state, a temperature gradient $\Delta T/\Delta z$ along the transport direction
$z$ can be extracted and the thermal conductivity $\kappa$ is calculated by applying Fourier's law of heat conduction,
\begin{equation}
\kappa=\frac{J}{\left|\Delta T/\Delta z\right|}.\label{eq:Fouriers law}
\end{equation}
Here we chose to divide the simulation box into eight slabs of equal width along the $z-$direction.
This will lead to at least $\sim 1500$ atoms per slab. Note that when dealing with PAA-PAP blends special
care needs to be taken in choosing the slab width because of the phase separation.
Velocity swapping was performed between the slowest atom in the center slab and the fastest atom in the
slab at the cell boundary. This swapping was performed every 20 fs with $\Delta t = 0.2$ fs.
After an initial steady-state equilibration for $10^{6}$ time-steps (see Figure \ref{fig:mp}(a)), the heat flux was computed over
a simulation time of $5\cdot10^{5}$ time-steps. Finally, a linear fit of the temperature profile as a
function of slab index was used to calculate $\kappa$ by means of Eq. (\ref{eq:Fouriers law}), as shown in Figure \ref{fig:mp}(b).

We have also attempted to calculate $\kappa$ using the equilibrium Kubo-Green method in LAMMPS \cite{greenkubo},
which, however, overestimates $\kappa$ by about an order of magnitude.
This can be attributed to the heat-flux autocorrelation function routine of the LAMMPS code that
only considers pair-wise interactions. Systems with many-body interactions may, therefore,
lead to problems in $\kappa$ calculations. More specifically, Kubo-Green should give the same
$\kappa$ values as in a non-equilibrium method. This, however, also require to properly
accounting angular and dihedral interactions in the all-atom force fields for the heat-flux calculations \cite{ohara08jcp}.
This was also identified earlier for the simulations of carbon based materials \cite{davideKG}.

\section{Results and discussions}
\label{sec:res}

\subsection{Morphology of polymer blends in the melt state}

\begin{figure*}[ptb]
\includegraphics[width=0.9\textwidth,angle=0]{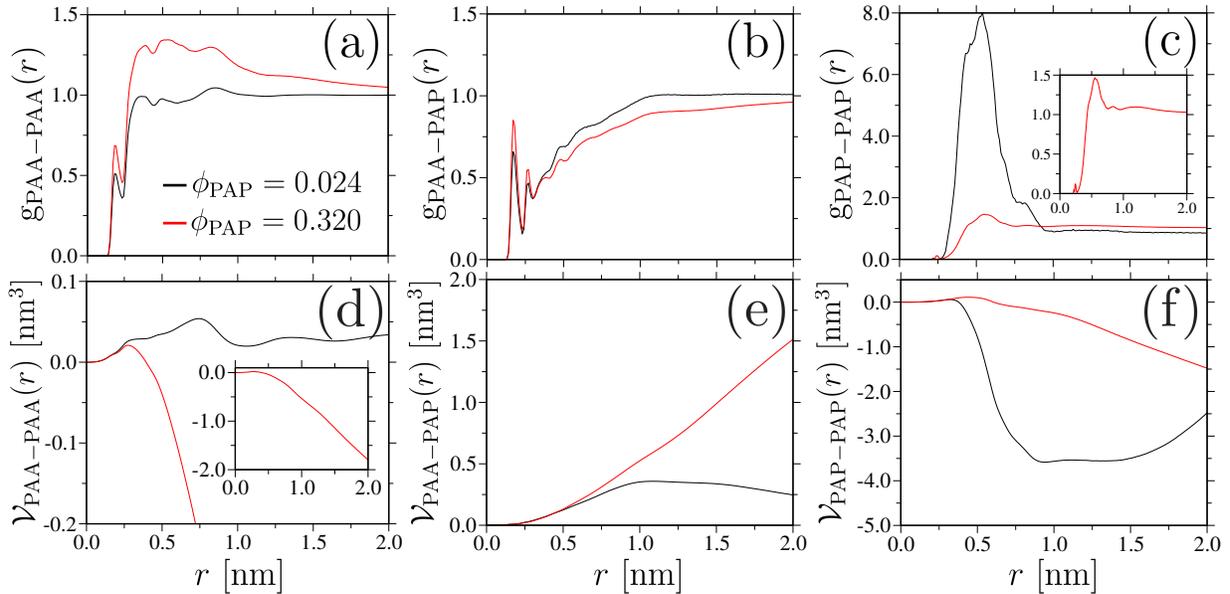}
        \caption{Radial distribution functions ${\rm g}_{\rm ij}(r)$ between three different monomeric pairs
        in polymer blends consisting of poly({\it N}-acrylyol piperidine) (PAP) and poly(acrylic acid) (PAA): (a)
        PAA-PAA, (b) PAA-PAP, and (c) PAP-PAP. Results are shown for two different PAP monomer molar fractions
        $\phi_{\rm PAP}$ under their melt states at a temperature of T$=600$ K and ambient pressure.
        Number of repeat units of PAP and PAA was chosen to be 3 and 30, respectively.
        For the calculation of ${\rm g}_{\rm ij}(r)$, we only consider oxygens and hydrogen of PAA and oxygen and
        nitrogen of PAP, as highlighted by blue species in Figure \ref{fig:schem}. Inset in part (c) is the enlarged view
        of ${\rm g}_{\rm PAP-PAP}(r)$ for $\phi_{\rm PAP} = 0.320$ highlighting first peak height. Parts (d-f) shows the
        cumulative integral of the second virial coefficient $\mathcal{V}_{ij} (r)$ between solution components.
        Inset in part (d) highlights the diverging $\mathcal{V}_{\rm PAA-PAA} (r)$ for $\phi_{\rm PAP} = 0.320$.
\label{fig:rdf}}
\end{figure*}

We start our discussion by investigating the molecular level morphologies of PAA-PAP blends at T = 600 K.
For this purpose, we calculate pair correlation functions ${\rm g}_{ij} (r)$ between different solution components.
Because the properties of these systems are dictated by hydrogen bonding, ${\rm g}_{ij}(r)$ are calculated
only between oxygen and hydrogen of PAA and oxygen and nitrogen of PAP, see highlighted blue components in Figure \ref{fig:schem}.
In Figures \ref{fig:rdf}(a-c), we show ${\rm g}_{ij} (r)$ between different monomeric species for two different $\phi_{\rm PAP}$.
A closer look at the data for $\phi_{\rm PAP} = 0.024$ (black curves in Figures \ref{fig:rdf}(a-c))
reveals two important length scales: (a) The most prominent fluid structure is observed for $r \le 1$ nm represented by the
correlation peaks. (b) ${\rm g}_{ij} (r) \to 1$ for $r \ge 1.5$ nm, which is the typical
correlation length in the hydrogen bonded systems \cite{mukherji13mac}. Furthermore, the long tail decay, as observed for $\phi_{\rm PAP} = 0.320$
(red curves in Figures \ref{fig:rdf}(a-c)), indicates that the system is at the onset of phase separation.

The pair correlation function ${\rm g}_{\rm ij}(r)$ not only gives information about the pairwise solution structure, it also provides
information about solution thermodynamics via the second virial
coefficient,
\begin{eqnarray}
        \mathcal{V}_{ij}(r) &=& 2\pi \int_0^{\infty} \left[1 - e^{-{\rm V}_{ij}(r)/k_{\rm B}T} \right] r^2 dr \nonumber\\
                            &=& 2\pi \int_0^{\infty} \left[1 - {\rm g}_{ij}(r) \right] r^2 dr.
\end{eqnarray}
This assumes $V_{ij}(r) = - k_{\rm B}T \ln\left[{\rm g}_{ij}(r)\right]$ \cite{originbi1}.
In polymer science, $\mathcal{V}_{ij}$ is also known as the excluded volume and is defined by the plateau value of the
cumulative integral ${\mathcal {V}}_{ij}(r)$ for $r$ value above the correlation length.
For example, the interaction between $i$ and $j$ is repulsive when $\mathcal{V}_{ij} > 0$ and attractive when $\mathcal{V}_{ij} < 0$.
When $\mathcal{V}_{ij} = 0$, the long range energetic attraction gets exactly canceled by the short range entropic repulsion, which is also
known as the ``so called" $\Theta-$point (or a critical point) that is dictated by large diverging fluctuations.
Note that the convergence of ${\mathcal {V}}_{ij}(r)$ for large $r$ values suffers from severe system size effects, especially
for multi-component solutions \cite{mukherji13mac}. Moreover, in this study we have chosen system sizes to be
large enough to avoid system size effects.

Figures \ref{fig:rdf}(d-f) show ${\mathcal {V}}_{ij}(r)$ for three different pairs of PAA-PAP blends.
For $\phi_{\rm PAP} = 0.024$ it can be appreciated that both PAA-PAA and PAA-PAP interactions are weak,
while the PAP-PAP interaction is highly attractive as indicated by large negative value of ${\mathcal {V}}_{\rm PAP-PAP}(r)$.
Almost equal preference for the interactions between PAA-PAA and PAA-PAP are not
surprising, given that these species are hydrogen bonded. Moreover,
the dominant contribution of the PAP-PAP interaction comes from the van der Waals
interaction between the hydrophobic side ring of PAP (highlighted by red in Figure \ref{fig:schem}).
We also want to emphasize that even when van der Waals interaction between two individual particles is
rather weak (i.e., less than $k_{\rm B}T$), collectively they may result in
several $k_{\rm B}T$ of interaction strength, as seen here for the PAP-PAP coordination.
Furthermore, a diverging ${\mathcal {V}}_{\rm PAA-PAP}(r)$ for $\phi_{\rm PAP} = 0.320$ further indicates
phase separation in PAA-PAP blend.

\begin{figure*}[ptb]
\includegraphics[width=0.9\textwidth,angle=0]{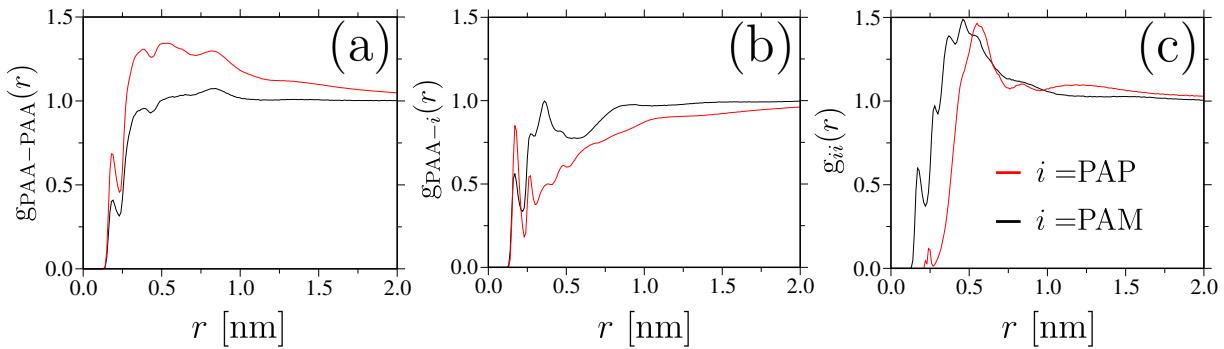}
        \caption{Radial distribution function ${\rm g}_{\rm ij}(r)$ between three different monomeric pairs
        in polymer blends for two different systems. One system is a blend of
        trimer of poly({\it N}-acrylyol piperidine) (PAP) is blended in with poly(acrylic acid) (PAA) with
        length $N_{\ell}^{\rm PAA} = 30$ (red curves) and the second systems is a mixture of PAA of same length with
        trimer of poly(acrylamide) (PAM) (black curves). The data is shown for 0.320 monomer mole fraction of
        shorter species, T$=600$ K and ambient pressure. Parts (a-c) show$-$ PAA-PAA, PAA-$i$ and $i-i$ structures,
        with the unit $i$ can either be PAP or PAM as specified in the legend.
\label{fig:rdf2}}
\end{figure*}

Solution processing of polymers with distinct nanoscopic interfaces, as in the case of phase separation, and
their use for tunable thermal, mechanical, optical and/or rheological properties is always a paramount challenge.
Therefore, it is desirable to have better miscible systems
for advanced applications. In this context, since the phase separation in a PAA-PAP blend is
dictated by interactions between the hydrophobic side groups of PAP, one possibility
to improve solubility of a blend might be to remove the side carbon ring in a PAP monomer structure.
Here PAM may serve as an ideal candidate (see Figure \ref{fig:schem}(c)).
PAM is an easy replacement because it has a similar monomer structure as PAP without
the carbon ring. Additionally, PAM is a water soluble polymer \cite{oliveira17jcp}, unlike PAP \cite{cahill16mac} that is
hydrophobic. The added advantage of using PAM arises from more possibility of hydrogen bonds in comparison to PAP, thus
forming stronger contacts between two particles.

In Figure \ref{fig:rdf2} we show a component-wise ${\rm g}_{ij}(r)$ for two different blends. It can be appreciated that the
system shows better tail convergence  for PAA-PAM systems in comparison to PAA-PAP blends, i.e., ${\rm g}_{ij}(r)=1$ around
the correlation length of 1.5 nm. This indicates a much better solubility in the system, as expected by the structural
tuning of the monomer units discussed in the preceding paragraph.
\begin{figure}[ptb]
\includegraphics[width=0.46\textwidth,angle=0]{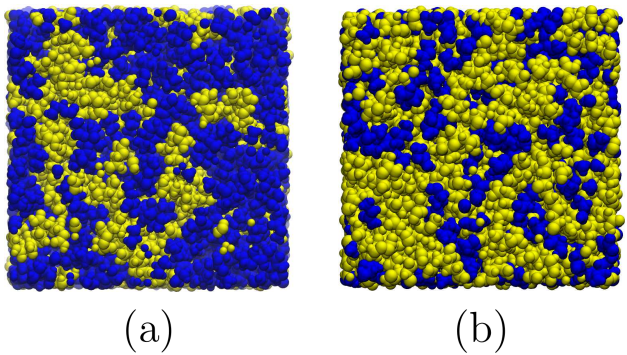}
        \caption{Space filling representation of the simulation snapshots for PAA-PAP blend part(a) and PAA-PAM blend part(b).
        The data is shown for 0.320 monomer mole fraction of shorter species, T$=600$ K and ambient pressure. Yellow spheres
        represent PAA atoms and blue spheres show PAP atoms (panel (a)) and  PAM atoms (panel (b)).
\label{fig:snap}}
\end{figure}
An illustration of molecular level morphologies in two blends are shown in Figure \ref{fig:snap}. It is evident that
PAA-PAM is more homogeneous, while PAA-PAP shows distinct islands.

Having discussed morphologies of smart polymer blends, we now move to understand the correlation between morphologies
and $\kappa$ in the dry states of these systems.

\subsection{Hydrogen bonding and thermal transport in polymer blends below the glass transition temperature}

\begin{figure}[ptb]
\includegraphics[width=0.4\textwidth,angle=0]{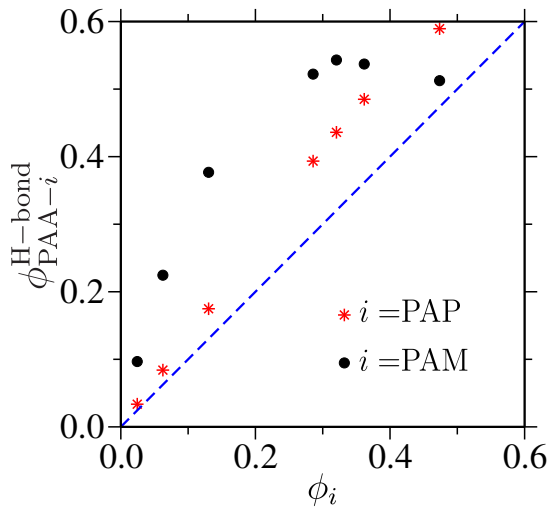}
\caption{Fraction of hydrogen bond $\phi^{\rm H-bond}_{{\rm PAA}-i}$ between cross components of two polymer
        blends (PAA-PAP and PAA-PAM blends) as a function of monomer mole fraction $\phi_i$, with $i$ can either be
        PAP or PAM. Blue dashed line is a linear interpolation between two values of $\phi_i$ with unit slope and
        zero intercept. The data is shown for the dry state of polymer blends for T$=300$ K and ambient pressure.
\label{fig:hbond}}
\end{figure}

Hydrogen bonding (H$-$bond) is an important molecular level interaction in these smart materials \cite{cohen10natmat,mukherji14natcom,winnik15review,hoogenboom,mukherji16sm}.
Therefore, we now investigate possible H$-$bonds between different solution species.
In Figure \ref{fig:hbond}, we show the fraction of H$-$bonds $\phi^{\rm H-bond}_{{\rm PAA}-i}$ between PAA and the other species $i$,
which can be either PAP or PAM. In a nutshell, if $\phi^{\rm H-bond}_{{\rm PAA}-i}$ is above the blue line (i.e.,
linear extrapolation between two concentrations with unity slope and zero intercept), there is an excess of
H$-$bonds for a given monomer mole fraction $\phi_i$.

For the PAA-PAP systems (red stars in Figure \ref{fig:hbond}) it can be seen that
even when PAA and PAP phase separate, there is an excess concentration of H$-$bonds between PAA and PAP.
This is because both PAA and PAP form isolated islands having their
side groups dangling within the interface between two phase separated regions. Therefore,
they can still facilitate interfacial H$-$bonds acting as adhesive contacts between two islands.
Here it is also worth mentioning that the phase separation, as observed in PAA-PAP case, may not be a standard spinodal decomposition \cite{binder74prl}.
More specifically, the PAP oligomers are glued together by their hydrophobic contacts leading to the phase separation.
To better understand the thermodynamic origin of the phase separation in PAA-PAP blends a
more detailed investigation is needed, which is beyond the scope of the present study.
Furthermore, we expect these small islands to coarsen over longer simulation times
even if they are driven by weak surface tension. Moreover, our simulations already show
a clear signature that the PAA-PAP systems are at the onset of phase separation (see Figure \ref{fig:snap}).

\begin{figure}[h!]
\includegraphics[width=0.49\textwidth,angle=0]{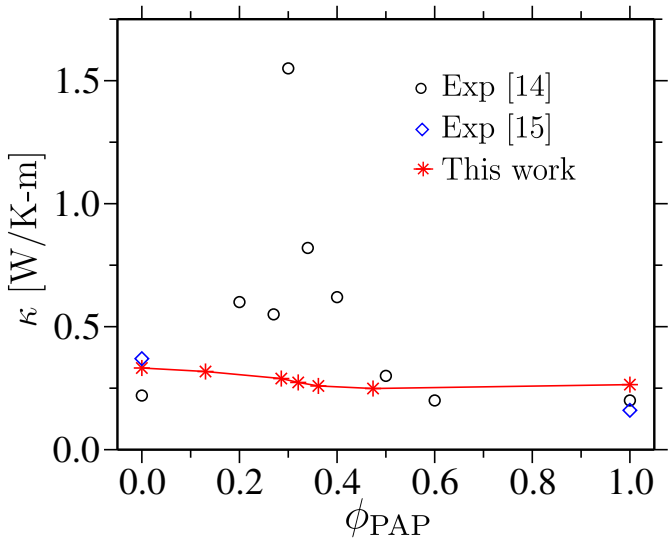}
        \caption{Thermal transport coefficient $\kappa$ for
        PAA-PAP blend as a function of PAP monomer mole fraction $\phi_{\rm PAP}$.
        The data is shown for the dry state (below the glass transition temperature) of polymer blends for
        temperature T$=300$ K and ambient pressure. A typical error of 10\% is estimated from five
        different $\kappa$ calculations using different random seeds during microcanonical simulations.
        For comparison, we have also included experimentally reported $\kappa$ values obtained for
        PAA-PAP blends \cite{pipe14nm} and the homopolymer data for pure PAA ($\phi_{\rm PAP}=0.0$)
        and for pure PAP ($\phi_{\rm PAP}=1.0$) \cite{cahill16mac}.
\label{fig:kappa}}
\end{figure}

Having two glued regions, as in the case of PAA-PAP blends, does not necessarily mean that one can also expect to have a variation
in $\kappa$ within the intermediate mixing ratios of $\phi_{\rm PAP}$. Instead the overall behavior
is expected to be dominated by $\kappa$ values of two individual components, with rather weak interfacial interactions.
Therefore, following the simple mixing rule, one should only expect a smooth interpolation of $\kappa$ between two pure phases of PAA and PAP.
Indeed, as shown by the simulation data in Figure \ref{fig:kappa} (red stars), $\kappa$ varies rather monotonically with $\phi_{\rm PAP}$.
It should also be noted that$-$ for the pure phases of PAA and PAP, our calculated values $\kappa \sim 0.32$ Wm$^{-1}$K$^{-1}$ (for $\phi_{\rm PAP} = 0.0$)
and $\kappa \sim 0.27$ Wm$^{-1}$K$^{-1}$ (for $\phi_{\rm PAP} = 1.0$) are in good agreement with the experimental data, see Figure \ref{fig:kappa} \cite{pipe14nm,cahill16mac}.
For example, one experiment reported $\kappa \sim 0.22$ Wm$^{-1}$K$^{-1}$ (for $\phi_{\rm PAP} = 0.0$) and $\kappa \sim 0.20$ Wm$^{-1}$K$^{-1}$ (for $\phi_{\rm PAP} = 1.0$) \cite{pipe14nm},
while another set of experimental data reported $\kappa \sim 0.37$ Wm$^{-1}$K$^{-1}$ (for $\phi_{\rm PAP} = 0.0$) and $\kappa \sim 0.16$ Wm$^{-1}$K$^{-1}$ (for $\phi_{\rm PAP} = 1.0$) \cite{cahill16mac}.
Furthermore, our simulation data for intermediate mixing ratios of $\phi_{\rm PAP}$ is in clear contradiction with one set of earlier
published experimental data \cite{pipe14nm} (see the data set corresponding to the black circles in Figure \ref{fig:kappa}),
while it is in agreement with another set of experimental observations \cite{cahill16mac}.

\begin{figure*}[ptb]
\includegraphics[width=0.97\textwidth,angle=0]{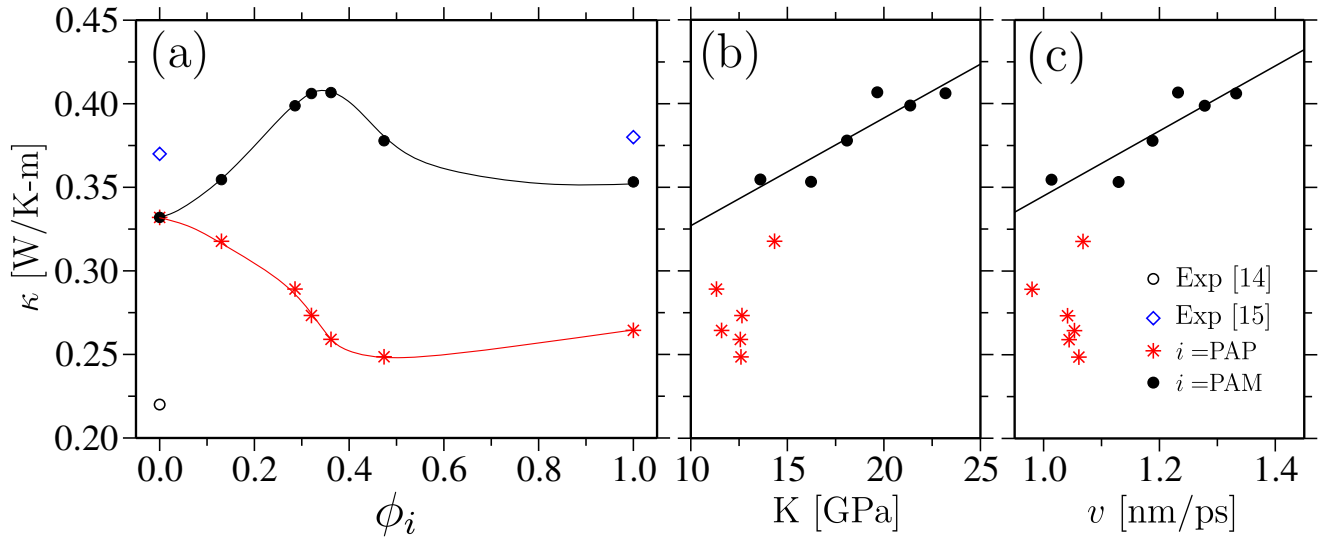}
        \caption{Part (a) shows comparative data of thermal transport coefficient $\kappa$ obtained from simulations for
        PAA-PAM and PAA-PAP blends as a function of monomer mole fraction $\phi_i$, with $i$ can either be
        PAP or PAP. The data is shown for the dry state (below glass transition temperature) of polymer blends for
        temperature T$=300$ K and ambient pressure. A typical error of 10\% is estimated from five
        different $\kappa$ calculations using different random seeds during microcanonical simulations.
        For comparison, we have also included experimentally reported $\kappa$ values obtained for
        pure PAA ($\phi_{\rm PAP}=0.0$) \cite{pipe14nm,cahill16mac} and pure PAM ($\phi_{\rm PAP}=1.0$) \cite{cahill16mac}.
        Lines are drawn to guide the eye.
        Parts (b-c) show $\kappa$ as a function of bulk modulus $K$ and sound velocity $v = \sqrt{K/\rho}$ calculated for both blends,
        where $\rho$ is the mass density. The black solid lines are linear fits to the PAA-PAM data.
\label{fig:kappa2}}
\end{figure*}

For the PAA-PAM system, we observe a higher $\phi^{\rm H-bond}_{{\rm PAA}-{\rm PAM}}$ (black filled circles in Figure \ref{fig:hbond}).
This excess is also coupled with an improvement in $\kappa$ for PAA-PAM blends in comparison to PAA-PAP systems, see
corresponding data with black filled circles in Figure \ref{fig:kappa2}(a).
The improvement of $\kappa$ for PAA-PAM as compared to PAA-PAP is not surprising, given that PAA and PAM
are fairly miscible because of preferential H$-$bonding between PAA and PAM (see Figure \ref{fig:rdf2} and Figure \ref{fig:hbond}).
Moreover, to further investigate the tunability of $\kappa$, we have also looked into the minimal thermal
conductivity model \cite{cahill16mac,einstein,cahill93prb}. Within this theory for amorphous polymer (as in our cases),
$\kappa$ relates directly to the materials stiffness, thus is also related to the glass transition temperature
$T_{\rm g}$ of amorphous systems \cite{pipe14nm} and the sound wave velocity $v$.
The higher the stiffness (or $v$), the larger the corresponding $\kappa$ \cite{kappaK}. In this context and
for the pure phases of PAA, PAP and PAM, we find that our calculated $\kappa$ values follow the
trend $\kappa^{\rm PAM} > \kappa^{\rm PAA} > \kappa^{\rm PAP}$ and are consistent
with $T_{\rm g}^{\rm PAM} > T_{\rm g}^{\rm PAA} > T_{\rm g}^{\rm PAP}$, see the supplementary Fig. 1 and Table I.
Furthermore, the sound velocity can be estimated using the Newton-Laplace equation $v = \sqrt{K/\rho}$,
where $\rho$ is the mass density and $K$ is the bulk modulus. Here, volume fluctuations are used to
calculate $K$ from NpT simulations using the expression
\begin{equation}
        K = k_{\rm B}T \frac {\left<V\right>} {\left<V^2\right> - \left<V\right>^2}.
\end{equation}
As expected from the theory \cite{cahill16mac,einstein,cahill93prb} and observed earlier in experiments \cite{cahill16mac},
our simulation data for PAA-PAM blends show that $\kappa$ is proportional to $K$ and $v$, see black symbols in Figures \ref{fig:kappa2}(b-c).
Additionally, the lack of correlation between $\kappa$ and $K$ (or $v$) for PAA-PAP is due to the phase separation in the systems,
see red symbols in Figures \ref{fig:kappa2}(b-c).

From the $K$ and $v$ values, we have also estimated the typical range of Debye temperatures $\Theta_{\rm D}$ using the expression in Ref. \cite{debyeglass}.
$\Theta_{\rm D}$ ranges between 200$-$250 K for different blends, while simulations are conducted at $T= 300$ K.
Therefore, quantum effects can be neglected \cite{solid}, thus classical molecular dynamics is an appropriate technique for these systems.

Figure \ref{fig:kappa2}(a) also shows that $\kappa$ for PAA-PAM systems first weakly increases and then
decreases again, attaining a maximum around $\phi_{\rm PAM} \sim 0.30$ (see black filled circles).
On the other hand, $\kappa$ for PAA-PAP systems decreases with $\phi_{\rm PAP}$ (see red stars).
Here the PAA-PAM H$-$bonds are preferred (over PAA-PAA and PAM-PAM H$-$bonds) because the maximum possible H$-$bonds between a PAA and a PAM monomer
is about four, which collectively can lead to more than $10 k_{\rm B}T$ energy per contact. On the other hand, two PAA or two PAM monomers
can maximally have one or two possible H$-$bonds between them, respectively \cite{oliveira17jcp}, thus leading to lesser contact
energy between same monomeric species.

\begin{figure}[ptb]
\includegraphics[width=0.49\textwidth,angle=0]{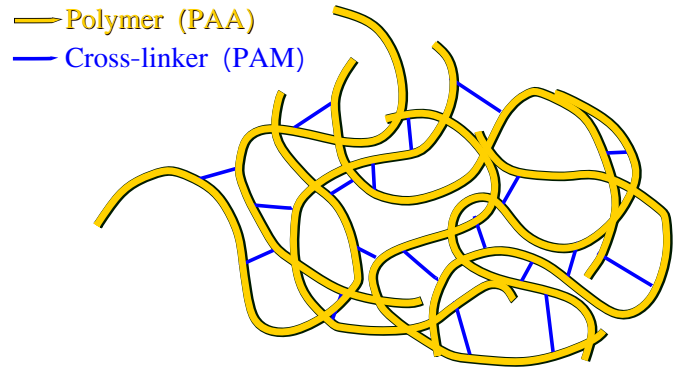}
        \caption{A schematic representation of short PAM oligomers (blue lines) forming bridges between
        PAA monomers (yellow lines).
\label{fig:schem2}}
\end{figure}

The strongly H$-$bonded contact between PAA and PAM can also explain the non-monotonous variation
of $\kappa$ with $\phi_{\rm PAM}$ (see black solid circles in Figure \ref{fig:kappa2}(a)).
In this context, we find that the short PAM oligomers act as
cross-linking bridges between two (or more) PAA monomers belonging to different polymers.
A simplified schematic of this bridging scenario is presented in Figure \ref{fig:schem2}.
Here the degree of cross linking is dictated by $\phi_{\rm PAM}$. The higher the $\phi_{\rm PAM}$ upto a
threshold concentration, the larger the bridging connectivity and
thus increased stiffness (as estimated from $K$) of materials.
This increased $K$ then leads to elevated $\kappa$. Furthermore, the maximum value of $\kappa$ is observed
around $\phi_{\rm PAM} \sim 0.3-0.4$. This is expected because when a small amount of PAM are blended in
the PAA material, each PAM oligomer will bind to more than one PAA monomer to reduce the binding free energy \cite{mukherji14natcom,mukherji17polsc}.
It should also be noted that the size of a PAM trimer is of the order of 0.75 nm,
which is also typically equivalent to 2$-$3 times the PAA monomer size (see Figure \ref{fig:schem}).
This length scale consistency also leads to almost perfect packing for PAA-PAM systems and thus forming
three dimensional cross-linking networks.
Moreover, when $\phi_{\rm PAM}$ is increased above a threshold value (for example $\phi_{\rm PAM} > 0.4-0.5$)
the effect is expected to be diluted because almost all PAA monomer will have at least one
PAM to bind. Therefore, $\kappa$ values will then be dominated by the pure phases of the individual polymers, see Figure \ref{fig:kappa2}(a).

Lastly we would also like to emphasize that even when PAA-PAP blends phase separate, leading to no noteceable change in
$\kappa$ with varying $\phi_{\rm PAP}$ (see Figure \ref{fig:kappa}), di-block copolymer architectures (consisting of PAA and PAP blocks) may lead to interesting
lamellar mesophases \cite{leibler,muser} and, therefore, providing another route towards the tunability of $\kappa$ in amorphous polymers.
In this context, layered superlattices (or thermal band gap materials) have been shown to exhibit interesting thermal behavior because
of their interfacial properties \cite{maldovan}. Here, however, it should also be noted that in lamellar phases of di-block copolymers,
unlike superlattices, phonon interference may not be expected to give a dominant contribution because the phonon mean free path is
vanishingly small in amorphous solids and polymers \cite{cahill16mac,cahill93prb}.

\section{Conclusions and outlook}
\label{sec:conc}

In this work, we have used molecular dynamics simulations to study thermal transport of asymmetric smart polymer blends
and its connection to underlying macromolecular morphologies. For this purpose, we investigate two experimentally relevant polymer blends. Our
structural analysis suggests that$-$ while a system of PAA-PAP blends are at the onset of phase separation,
a system of PAA and PAM is fairly well miscible with significant excess hydrogen bonded interaction between
cross species. The short PAM chains act as cross-linking bridges between monomers of
different PAA chains forming a three dimensional (hydrogen bonded) highly cross-linked smart polymer network,
thus increasing materials stiffness and improved thermal transport coefficient $\kappa$.
We want to emphasize that the absolute values of $\kappa$ calculated in our simulations
are within the experimental uncertainty and also consistent with the stiffness measurements.
A rather generic picture emerging of these results is that $\kappa$ may be tuned when
a system satisfies a few key conditions: miscibility, preferential hydrogen bonding, and
the formation of cross-linking networks. Although we have presented data for
PAA-PAM systems with improved $\kappa$, our simulation results indicate a rather generic design principle for plastic
materials with the improvement and tunability of $\kappa$.
To validate the protocol presented in this work,
more detailed experimental synthesis, characterization, and their thermal transport measurements are needed
on a broader spectrum of polymeric materials. Moreover, results presented here may serve as a
guiding principle for the operational understanding and functional design of advanced materials with
tunable properties.

\section{Acknowledgement}

D.M. thank Kurt Kremer and Carlos M. Marques for fruitful continual collaborations that lead to
the understanding of smart polymers presented here. We thank Derek Fujimoto for useful
discussions. We further acknowledge support from Compute Canada where simulations were performed.

\end{document}